\documentclass[aps,prl,twocolumn,showpacs,superscriptaddress]{revtex4-2} 
\usepackage{times}
\usepackage{graphicx} 
\usepackage{graphics}
\usepackage{dcolumn} 
\usepackage{bm} 
\usepackage{amssymb} 
\usepackage{siunitx} 
\usepackage{amsmath}
\usepackage{amssymb}
\usepackage{float}
\usepackage{epstopdf}
\usepackage[colorlinks = true]{hyperref}
\hypersetup{
colorlinks=true,citecolor=blue,urlcolor=blue,linkcolor=blue 
}

\begin{document}

\title[Article Title]{Boosted fusion gates above the percolation threshold for scalable graph-state generation}

\author{Yong-Peng Guo}
\altaffiliation{These authors contributed equally to this work.}
\affiliation{Hefei National Research Center for Physical Sciences at the Microscale and School of Physical Sciences,
University of Science and Technology of China, Hefei, Anhui 230026, China}
\affiliation{CAS Center for Excellence in Quantum Information and Quantum Physics,
University of Science and Technology of China, Hefei, Anhui 230026, China}
\affiliation{Hefei National Laboratory,
University of Science and Technology of China, Hefei, Anhui 230088, China}

\author{Geng-Yan Zou}
\altaffiliation{These authors contributed equally to this work.}
\affiliation{Hefei National Research Center for Physical Sciences at the Microscale and School of Physical Sciences,
University of Science and Technology of China, Hefei, Anhui 230026, China}
\affiliation{CAS Center for Excellence in Quantum Information and Quantum Physics,
University of Science and Technology of China, Hefei, Anhui 230026, China}
\affiliation{Hefei National Laboratory,
University of Science and Technology of China, Hefei, Anhui 230088, China}

\author{Xing Ding}
\altaffiliation{These authors contributed equally to this work.}
\affiliation{Hefei National Research Center for Physical Sciences at the Microscale and School of Physical Sciences,
University of Science and Technology of China, Hefei, Anhui 230026, China}
\affiliation{CAS Center for Excellence in Quantum Information and Quantum Physics,
University of Science and Technology of China, Hefei, Anhui 230026, China}
\affiliation{Hefei National Laboratory,
University of Science and Technology of China, Hefei, Anhui 230088, China}

\author{Qi-Hang Zhang}
\altaffiliation{These authors contributed equally to this work.}
\affiliation{Hefei National Research Center for Physical Sciences at the Microscale and School of Physical Sciences,
University of Science and Technology of China, Hefei, Anhui 230026, China}
\affiliation{CAS Center for Excellence in Quantum Information and Quantum Physics,
University of Science and Technology of China, Hefei, Anhui 230026, China}
\affiliation{Hefei National Laboratory,
University of Science and Technology of China, Hefei, Anhui 230088, China}

\author{Mo-Chi Xu}
\affiliation{Hefei National Research Center for Physical Sciences at the Microscale and School of Physical Sciences,
University of Science and Technology of China, Hefei, Anhui 230026, China}
\affiliation{CAS Center for Excellence in Quantum Information and Quantum Physics,
University of Science and Technology of China, Hefei, Anhui 230026, China}
\affiliation{Hefei National Laboratory,
University of Science and Technology of China, Hefei, Anhui 230088, China}

\author{Run-Ze Liu}
\affiliation{Hefei National Research Center for Physical Sciences at the Microscale and School of Physical Sciences,
University of Science and Technology of China, Hefei, Anhui 230026, China}
\affiliation{CAS Center for Excellence in Quantum Information and Quantum Physics,
University of Science and Technology of China, Hefei, Anhui 230026, China}
\affiliation{Hefei National Laboratory,
University of Science and Technology of China, Hefei, Anhui 230088, China}

\author{Jun-Yi Zhao}
\affiliation{Hefei National Research Center for Physical Sciences at the Microscale and School of Physical Sciences,
University of Science and Technology of China, Hefei, Anhui 230026, China}
\affiliation{CAS Center for Excellence in Quantum Information and Quantum Physics,
University of Science and Technology of China, Hefei, Anhui 230026, China}
\affiliation{Hefei National Laboratory,
University of Science and Technology of China, Hefei, Anhui 230088, China}

\author{Zhen-Xuan Ge}
\affiliation{Hefei National Research Center for Physical Sciences at the Microscale and School of Physical Sciences,
University of Science and Technology of China, Hefei, Anhui 230026, China}
\affiliation{CAS Center for Excellence in Quantum Information and Quantum Physics,
University of Science and Technology of China, Hefei, Anhui 230026, China}
\affiliation{Hefei National Laboratory,
University of Science and Technology of China, Hefei, Anhui 230088, China}

\author{Li-Chao Peng}
\affiliation{MIIT Key Laboratory of Complex-field Intelligent Exploration, School of Optics and Photonics, Beijing Institute of Technology, Beijing, 100081, China}

\author{Ke-Mi Xu}
\affiliation{MIIT Key Laboratory of Complex-field Intelligent Exploration, School of Optics and Photonics, Beijing Institute of Technology, Beijing, 100081, China}

\author{Yi-Yang Lou}
\affiliation{Hefei National Research Center for Physical Sciences at the Microscale and School of Physical Sciences,
University of Science and Technology of China, Hefei, Anhui 230026, China}
\affiliation{CAS Center for Excellence in Quantum Information and Quantum Physics,
University of Science and Technology of China, Hefei, Anhui 230026, China}
\affiliation{Hefei National Laboratory,
University of Science and Technology of China, Hefei, Anhui 230088, China}

\author{Zhen Ning}
\affiliation{Hefei National Research Center for Physical Sciences at the Microscale and School of Physical Sciences,
University of Science and Technology of China, Hefei, Anhui 230026, China}
\affiliation{CAS Center for Excellence in Quantum Information and Quantum Physics,
University of Science and Technology of China, Hefei, Anhui 230026, China}
\affiliation{Hefei National Laboratory,
University of Science and Technology of China, Hefei, Anhui 230088, China}

\author{Lin-Jun Wang}
\affiliation{Center for Micro and Nanoscale Research and Fabrication,
University of Science and Technology of China, Hefei, Anhui 230026, China}

\author{Hui Wang}
\affiliation{Hefei National Research Center for Physical Sciences at the Microscale and School of Physical Sciences,
University of Science and Technology of China, Hefei, Anhui 230026, China}
\affiliation{CAS Center for Excellence in Quantum Information and Quantum Physics,
University of Science and Technology of China, Hefei, Anhui 230026, China}
\affiliation{Hefei National Laboratory,
University of Science and Technology of China, Hefei, Anhui 230088, China}

\author{Yong-Heng Huo}
\affiliation{Hefei National Research Center for Physical Sciences at the Microscale and School of Physical Sciences,
University of Science and Technology of China, Hefei, Anhui 230026, China}
\affiliation{CAS Center for Excellence in Quantum Information and Quantum Physics,
University of Science and Technology of China, Hefei, Anhui 230026, China}
\affiliation{Hefei National Laboratory,
University of Science and Technology of China, Hefei, Anhui 230088, China}

\author{Yu-Ming He}
\affiliation{Hefei National Research Center for Physical Sciences at the Microscale and School of Physical Sciences,
University of Science and Technology of China, Hefei, Anhui 230026, China}
\affiliation{CAS Center for Excellence in Quantum Information and Quantum Physics,
University of Science and Technology of China, Hefei, Anhui 230026, China}
\affiliation{Hefei National Laboratory,
University of Science and Technology of China, Hefei, Anhui 230088, China}

\author{Chao-Yang Lu}
\affiliation{Hefei National Research Center for Physical Sciences at the Microscale and School of Physical Sciences,
University of Science and Technology of China, Hefei, Anhui 230026, China}
\affiliation{CAS Center for Excellence in Quantum Information and Quantum Physics,
University of Science and Technology of China, Hefei, Anhui 230026, China}
\affiliation{Hefei National Laboratory,
University of Science and Technology of China, Hefei, Anhui 230088, China}

\author{Jian-Wei Pan}
\affiliation{Hefei National Research Center for Physical Sciences at the Microscale and School of Physical Sciences,
University of Science and Technology of China, Hefei, Anhui 230026, China}
\affiliation{CAS Center for Excellence in Quantum Information and Quantum Physics,
University of Science and Technology of China, Hefei, Anhui 230026, China}
\affiliation{Hefei National Laboratory,
University of Science and Technology of China, Hefei, Anhui 230088, China}

\date{\today}

\begin{abstract}
Fusing small resource states into a larger, fully connected graph-state is essential for scalable photonic quantum computing. Theoretical analysis reveals that this can only be achieved when the success probability of the fusion gate surpasses a specific percolation threshold of 58.98\% by using three-photon GHZ states as resource states. However, such an implementation of a fusion gate has never been experimentally realized before. Here, we successfully demonstrate a boosted fusion gate with a theoretical success probability of 75\%, using deterministically generated auxiliary states. The success probability is experimentally measured to be 71.0(7)\%. We further demonstrate the effectiveness of the boosted fusion gate by fusing two Bell states with a fidelity of 67(2)\%. Our work paves a crucial path toward scalable linear optical quantum computing.
\end{abstract}

\maketitle

Quantum computing, with its potential to address computationally intractable problems beyond the reach of classical computation~\cite{aaronson2011computational,jz1,Google2019}, has witnessed significant advancements in both qubit quantity and quality, propelling unprecedented progress in quantum error correction~\cite{Lukin,GOOGLE}. Photonic systems have long coherence time, precise single-qubit operations, and room-temperature compatibility to quantum networks, presenting a promising avenue toward the realization of fault-tolerant quantum computation~\cite{Rudolph2005,duan2005efficient,varnava2006loss,kieling2007percolation,Rudolph2008,stace2009thresholds,fujii2010fault,li2010fault,gimeno2015three,li2015resource,auger2018fault,Mihir2019perolation,bartolucci2023fusion}.

One approach for photonic quantum computing is through Bell fusion mechanisms~\cite{Rudolph2005,duan2005efficient,kieling2007percolation,Rudolph2008,li2010fault,gimeno2015three,li2015resource,auger2018fault,Mihir2019perolation,bartolucci2023fusion}. In this model, the required large-scale graph-states~\cite{Nielson2004} can be prepared by fusing smaller resource states. This is enabled by percolation theory, which provides scalable methods for generating fully connected graph-states via probabilistic fusion operations. This theory predicts a critical fusion probability, $p_c$, above which the size of the generated connected graph-states scales linearly with the number of resource states, rendering this approach suitable for linear optical quantum computing~\cite{kieling2007percolation}. One minimal resource states are three-photon Greenberger-Horne-Zeilinger (GHZ) states, which were experimentally realized very recently~\cite{chen2023heralded,cao2023photonic,maring2023general}. A key remaining challenge is the realization of high-efficiency fusion gates exceeding the requisite threshold.

Recent advances in percolation theory have shown that exceeding a critical percolation threshold, initially estimated at approximately $\sim$62.5\%~\cite{gimeno2015three} and subsequently improved to 58.98\%~\cite{Mihir2019perolation}, enables the construction of scalable infinite-dimensional graph-states from three-photon GHZ states~\cite{Rudolph06}. 
Although a fusion gate with a success probability of 57.9\% has recently been realized~\cite{bayerbach2023bell}, unfortunately, it still falls short of the threshold required for scalable linear optical quantum computing.

\begin{figure}[tb]%
\centering
\includegraphics[width=0.4\textwidth]{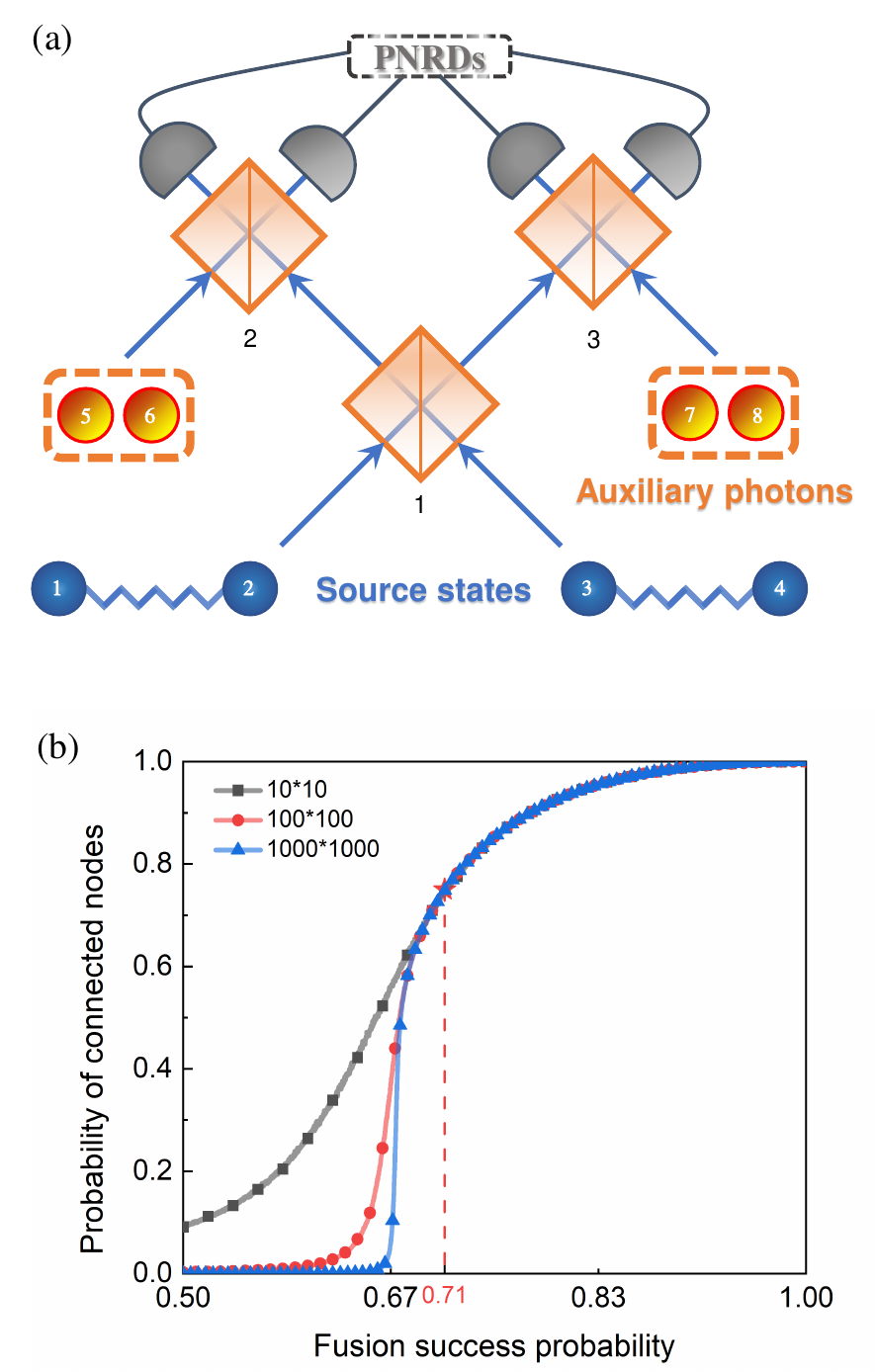}
\caption{The boosted type-$\mathrm{\uppercase\expandafter{\romannumeral2}}$ fusion gate. (a) The general case of a boosted type-$\mathrm{\uppercase\expandafter{\romannumeral2}}$ fusion gate. (b) The probability of connected nodes when building 2D-cluster states of certain sizes. 3-photon GHZ states are used as source states, and in this circumstance, the percolation threshold is 0.672. 0.71 is the experimental performance of fusion gates using 4 auxiliary photons.
}\label{fig1}
\end{figure}

In this work, we report a boosted fusion gate surpassing the percolation threshold of 58.98\%, with an experimentally determined success probability of 71.0(7)\%. This is realized by fusing two Bell states with four auxiliary photons~\cite{grice2011arbitrarily,0.75}, as shown in Fig.~\ref{fig1}(a). Measurement of the fusion process yields a fidelity of 67(2)\%, providing the first direct experimental evidence of the effectiveness of a boosted fusion gate. 

To assess the advantage of our high-performance fusion gates in the generation of large-scale graph-states, simulations are performed using a modified Newman-Ziff algorithm~\cite{löbl2023efficient}. In our simulation, 3-photon GHZ states are utilized as resource states to build 2D-cluster states with sizes of 10$\times$10, 100$\times$100, and 1000$\times$1000. The numerical result is shown in Fig.~\ref{fig1}(b), where there is a specific threshold of 0.672~\cite{Mihir2019perolation} when photon loss is ignored. Below this threshold, a large cluster state cannot be connected efficiently. Near this threshold, 3-photon GHZ resource states can be quickly fused and renormalized to form a 2D-cluster state. Moreover, larger connected states will be allowed with higher success probabilities of fusion gates. The experimentally demonstrated success probability of our fusion gate is indicated by the red dotted line, highlighting its advantage in generating larger graph-states.

\begin{figure*}[tb]%
\centering
\includegraphics[width=1\textwidth]{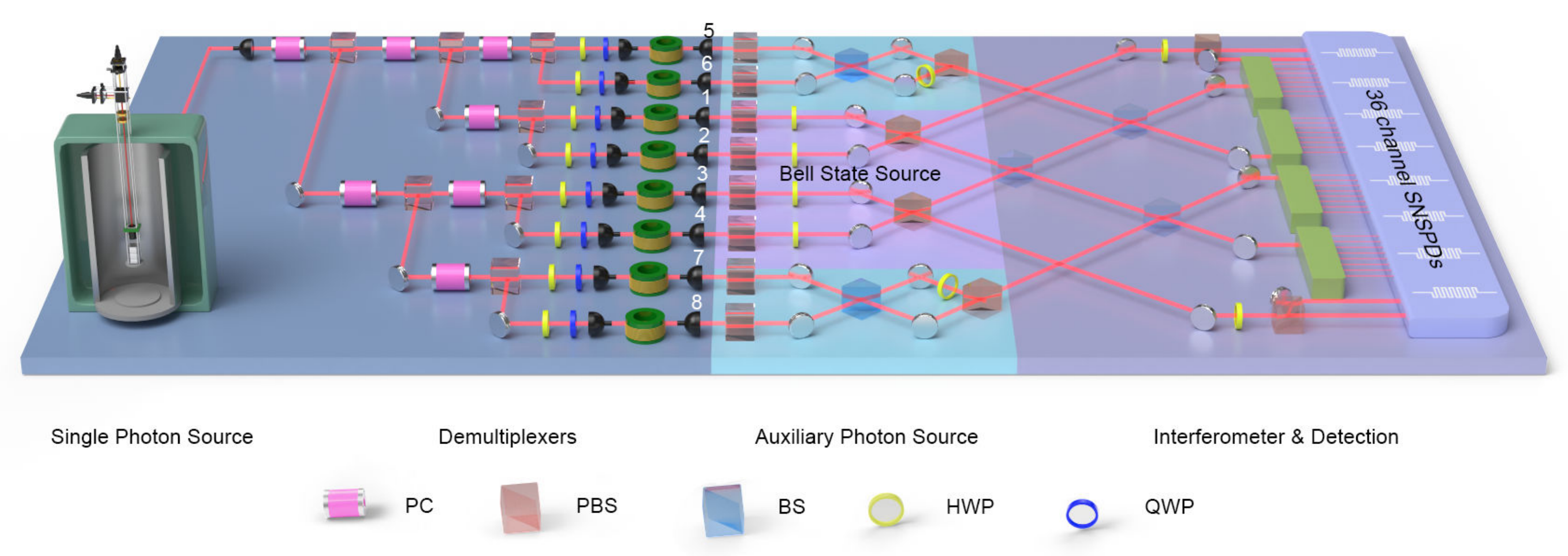}
\caption{Experimental setup of the boosted type-$\mathrm{\uppercase\expandafter{\romannumeral2}}$ fusion gate. The single photons are demultiplexed by 7 Pockels cells into 8 spatial modes. Two Bell states $|\Phi^+\rangle_{12}$ and $|\Phi^+\rangle_{34}$ are post-selected from photons 1, 2 and 3, 4. Auxiliary states $|\gamma\rangle_{56}$ and $|\gamma\rangle_{78}$ are prepared by HOM interference. To demonstrate a complete fusion operation, an enhanced BSM is performed on photons 2 and 3, while polarization analysis is conducted on photons 1 and 4 to prove the effectiveness. All effective eight-photon coincidence events that occur are recorded with pseudo-photon-number-resolving detectors constructed by 36-channel superconducting nanowire single-photon detectors.
}\label{fig2}
\end{figure*}

The single photons required are generated through the resonant excitation of a self-assembled InGaAs quantum dot within a tunable open cavity~\cite{ding2023highefficiency}. This single-photon source demonstrates a single-photon purity of 0.9795, corrected photon indistinguishability of 0.9856, and an overall system efficiency of 0.712 as outlined in Ref.~\cite{ding2023highefficiency}. The outstanding photon indistinguishability and system efficiency of this single-photon source guarantee both high fidelity and a substantial coincident rate for the fusion gate in this study.

\begin{figure*}[tb]%
\centering
\includegraphics[width=1.0\textwidth]{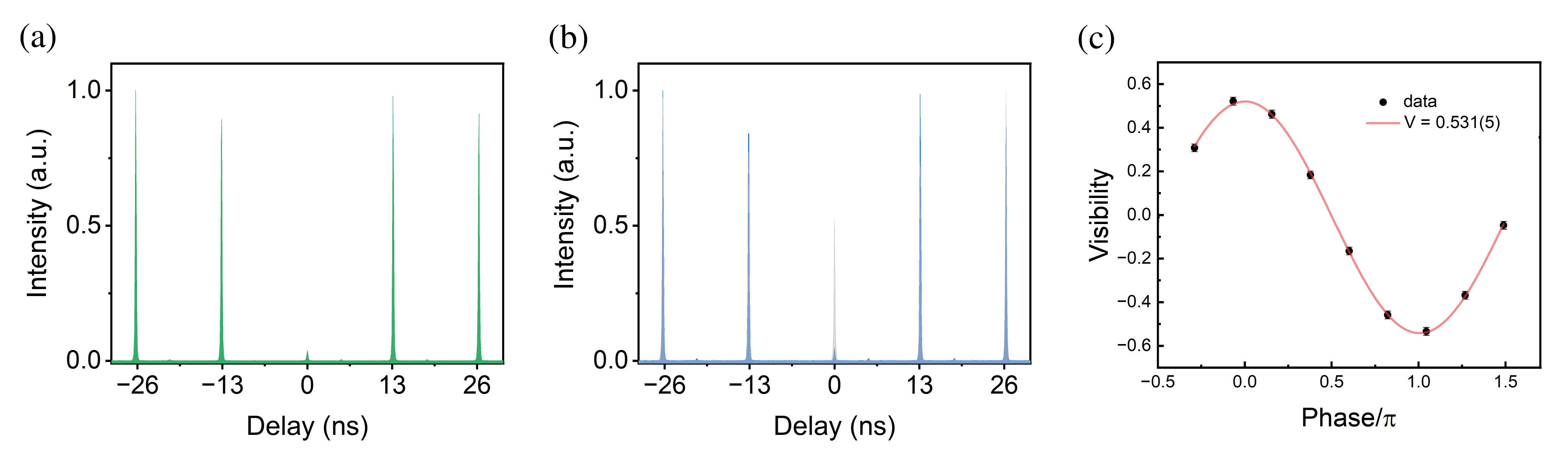}
\caption{Characterization of the experiment components. (a) Purity of the single-photon source after a 5-hour data collection. The second-order correlation is measured using the HBT setup \cite{HBT}, indicating a single photon purity of \strut $g^{(2)}(0)=0.0414(5)$. (b) The quality of auxiliary states. The two-photon correlations of photons 5 and 6 indicate a raw Hong-Ou-Mandel visibility of 0.9076(6). (c) The visibility of Bell state $|\Psi^-\rangle_{14}$ when traditional fusion gates are performed (without auxiliary photons). The off-diagonal element of the Bell state density matrix indicates a fidelity of 0.676(25).
}\label{fig3}
\end{figure*}

Subsequently, the single photon stream is demultiplexed into eight distinct spatial modes using a series of active switches~\cite{wang_2017_highefficiency}. Illustrated in Fig.~\ref{fig2}, these switches contain Pockels Cells (PCs), polarizing beam splitters (PBSs), waveplates, and optical fibers. In this configuration, each PC and PBS pair acts as an optical switch, achieving an extinction ratio exceeding 300:1. Synchronized to the pulsed laser and operating at a repetition rate of 710 kHz, each spatial mode receives a sequence of 10 single-photon pulses per cycle. Notably, the largest temporal separation between two demultiplexed single photons reaches an interval of $\sim$2.6 $\mu$s, where the photon indistinguishability remains as high as 0.959~\cite{ding2023highefficiency}.

The demultiplexed single photons are labeled as 1 to 8, with photons (1 - 4) serving as the resource photons to generate required Bell states and photons (5 - 8) acting as auxiliary photons to boost the success probability of the fusion gate. First, we prepare the Bell state $|\Phi^+\rangle_{12}$ ($|\Phi^+\rangle_{34}$) by sending $|+\rangle$ - polarized single photons 1 and 2 (3 and 4) into a PBS~\cite{Rudolph2005}, where $|+\rangle=(|H\rangle+|V\rangle)/\sqrt{2}$. Meanwhile, the auxiliary photons 5 and 6 (7 and 8) are prepared in a two-photon N00N state $|\gamma\rangle_{56}=\frac{1}{\sqrt{2}} (|2_{H}0_{V}\rangle_{56}+|0_{H}2_{V}\rangle_{56})$ ($|\gamma\rangle_{78}=\frac{1}{\sqrt{2}} (|2_{H}0_{V}\rangle_{78}+|0_{H}2_{V}\rangle_{78})$) based on the well-known Hong-Ou-Mandel (HOM) effect. As an example, two identical photons 5 and 6 are sent through a 50:50 beam splitter (BS), resulting in a two-photon N00N state $(|2_a0_b\rangle+|0_a2_b\rangle)/\sqrt{2}$, with $a, b$ denoting the two output ports of the beam splitter. Subsequently, with a $45^\circ$ half-wave plate (HWP) in port $b$ and a PBS, the target two-photon state ($|\gamma\rangle_{56}$) is achieved at the one port of the PBS.

To confirm the successful generation of auxiliary states $|\gamma\rangle_{56}$ and $|\gamma\rangle_{78}$, the HOM visibility of photons (5, 6), and photons (7, 8) are evaluated in our experiment, respectively. Benefited from the high purity of single photons (as shown in Fig.~\ref{fig3}(a)), the measured raw HOM visibility of photons (5, 6) and (7, 8) are 0.9076(6) and 0.9050(6) (as shown in Fig.~\ref{fig3}(b)), corresponding to a corrected visibility larger than 98\%. Additionally, the quality of prepared Bell states $|\Phi^+\rangle_{12}$ and $|\Phi^+\rangle_{34}$ are tested by fusing photons 2 and 3 on BS1~\cite{Rudolph2005}, which merges $|\Phi^+\rangle_{12}$ and $|\Phi^+\rangle_{34}$ into a new two-photon Bell state involving photon 1 and 4. To this end, the phase of one of the photons (photon 2) is changed from 0 to 2.2$\pi$, all measured Bell state visibilities are listed in Fig.~\ref{fig3}(c), where the largest (smallest) visibility at phase 0 ($\pi$) is 0.521 (-0.541), yielding a two-photon entanglement fidelity of 0.676(25). (see supplementary information for more details).

The core of our boosted type-$\mathrm{\uppercase\expandafter{\romannumeral2}}$ fusion operation is carried out following the enhanced Bell-state measurement (BSM) protocol~\cite{0.75}. This BSM procedure is organized as illustrated in Fig.~\ref{fig1}(a), using three 50: 50 BSs. Similar to a conventional BSM, photons 2 and 3 are directed into BS1 to distinguish $|\Psi^{+}\rangle_{23}$ and $|\Psi^{-}\rangle_{23}$. However, if photons 2 and 3 are in the states $|\Phi^{\pm}\rangle_{23}$, they remain indistinguishable at this stage. In such cases, both photons exit through the same port of the BS1 simultaneously, and the Bell states are subsequently distinguished by interacting photons (2, 3) with a two-photon auxiliary state on BS2 or BS3. For example, when photons (2, 3) interact with $|\gamma\rangle_{56}$ on BS2, four photons (2, 3, 5, 6) exit from BS2 and the other two auxiliary photons (7, 8) exit from BS3. The four-photon output state of BS2 corresponds to the theoretical predictions presented in~\cite{0.75}
\begin{equation}
\begin{aligned}
\label{Phi+}
|\Phi^{+}\rangle_{23} |\gamma\rangle_{56} \rightarrow& \frac18[-\sqrt6(|4000\rangle+|0040\rangle+|0400\rangle+|0004\rangle)\\
&-2(|2020\rangle+|0202\rangle)]+\frac14(-|2200\rangle+|2002\rangle\\
&+|0220\rangle -|0022\rangle-2|1111\rangle)
\end{aligned}
\end{equation}

\begin{equation}
\begin{aligned}
\label{Phi-}
|\Phi^{-}\rangle_{23} |\gamma\rangle_{56} \rightarrow& \frac18[-\sqrt6(|4000\rangle+|0040\rangle-|0400\rangle-|0004\rangle)\\
&-2(|2020\rangle-|0202\rangle)]+\frac{i\sqrt2}{4}(|2101\rangle-|1210\rangle\\
&+|1012\rangle-|0121\rangle)
\end{aligned}
\end{equation}

Each term in \eqref{Phi+} and \eqref{Phi-} represents a specific path and polarization distribution for the four output photons. For example, $|4000\rangle$ denotes that four horizontally polarized photons exit through a single port of BS2. We distinguish $|\Phi^+\rangle_{23}$ and $|\Phi^-\rangle_{23}$ by identifying their distinct output distributions in \eqref{Phi+} and \eqref{Phi-} with a 50\% success probability. Consequently, the theoretical success probability of a boosted type-$\mathrm{\uppercase\expandafter{\romannumeral2}}$ fusion gate, averaged across four Bell states, is 75\%.

\begin{figure*}[tb]%
\centering
\includegraphics[width=0.95\textwidth]{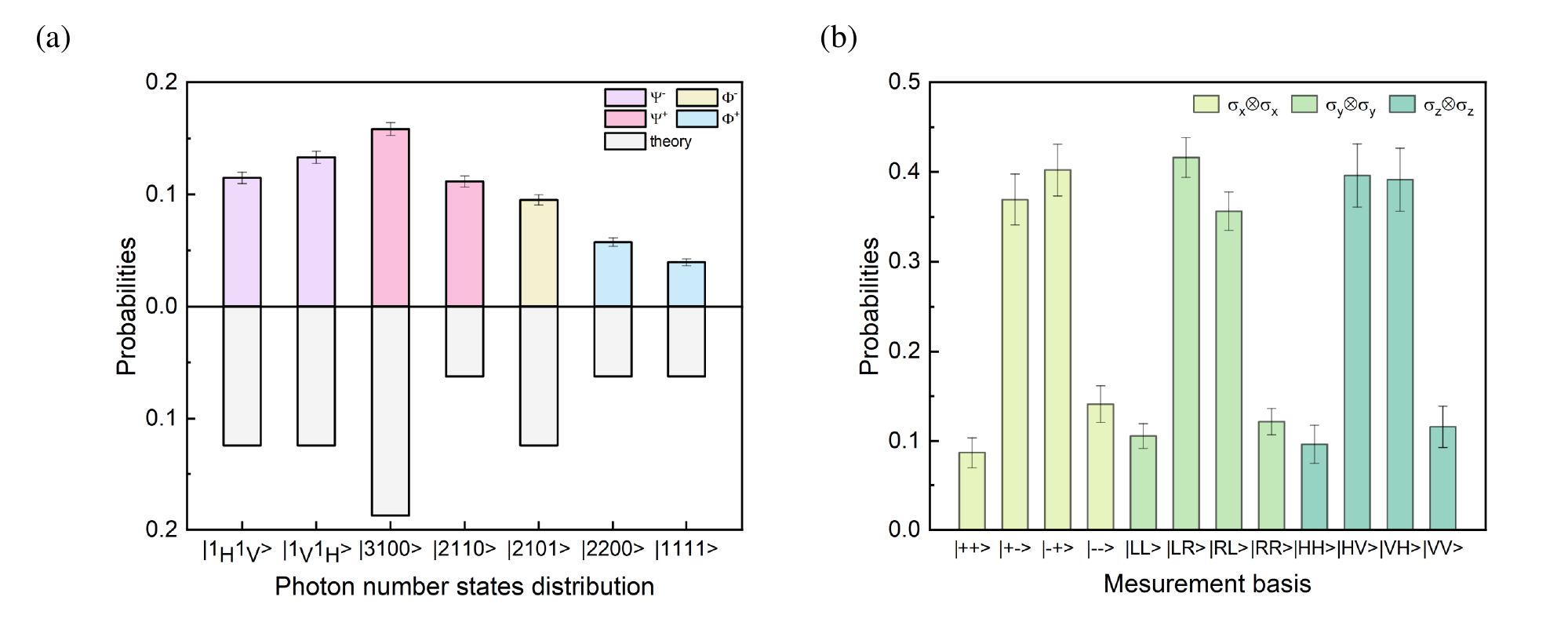}
\caption{Results of the boosted type-$\mathrm{\uppercase\expandafter{\romannumeral2}}$ fusion gate. (a) Detection probabilities of successful Bell state discrimination: Each colorful bar represents the detection probability of a certain photon distribution, which indicates successful discrimination of one Bell state. The probabilities are corrected by normalization factors.
The measured success probability of our fusion gate is 71.0(7)\%. The measured distributions are compared to theoretical ones represented by gray bars with a total theoretical success probability of 75\%. (b) Photon 1, 4 entanglement fidelity measurement: The bars show the results of measuring photon 1, 4 at the basis of $+-, RL, HV$ when they are fused to be $\vert{\Psi^-}\rangle_{14}$). The coincidence counts are corrected and the two-photon entanglement fidelity is calculated to be 67(2)\%.
}\label{fig4}
\end{figure*}

As specified in \eqref{Phi+} and \eqref{Phi-}, to analyze the output state of the interferometer, pseudo-photon-number-resolving detectors (PPNRDs) capable of distinguishing up to four photons are necessary. Photon polarization is first measured using a PBS, and each output mode of the PBS is further divided into four distinct modes using a 1$\times$4 multi-port BS for photon-number resolution. A 32-channel superconducting nanowire single-photon detector (SNSPD) is used for detecting photons 2 - 3 and 5 - 8. In addition, four more channels are allocated for the polarization projection measurement of photons 1 and 4. All valid eight-fold coincidence events are then recorded by a programmable multi-channel coincidence unit. An eight-fold coincidence rate of approximately 3 Hz is observed, corresponding to an end-to-end efficiency of $16\%$ for individual photons. Given an average detector efficiency of $72\%$, an overall transmission efficiency of $31\%$ can be estimated from the source to the output of the interference stage.

The detection probabilities of specific photon distributions are calculated to evaluate the success probability of our fusion gate, as shown in Fig.~\ref{fig4}(a). Each bar corresponds to a group of distributions representing successful detections of a particular Bell state. Since we can distinguish $|\Psi^-\rangle_{23}$ without an auxiliary state, the bars labeled $|1_H1_V\rangle$ and $|1_V1_H\rangle$ indicate the polarization distribution of photons (2, 3) at two output ports of BS1. Similarly, other bars represent the unique terms of the four-photon output state in \eqref{Phi+} and \eqref{Phi-}. Each measured probability is compared with its theoretical value after applying a normalization factor. The factor is determined based on the unequal detection probabilities of different multi-photon distributions. In our experiment, these normalization factors are calibrated at the start of data collection using a large dataset of coincident counts, ensuring minimal error in these values. Additional details of the normalization factors are provided in the supplementary material.

The corrected probabilities for detecting the states $|\Psi^-\rangle_{23}$, $|\Psi^+\rangle_{23}$, $|\Phi^-\rangle_{23}$, and $|\Phi^+\rangle_{14}$ are 0.248(5), 0.270(6), 0.095(5), and 0.097(4), respectively, yielding a total probability of successful Bell-states discrimination of 71.0(7)\%. The slight reduction from the theoretical probability of 75\% is attributed to the non-unity indistinguishability of photons, which can cause some successful events to result in unexpected photon-number distributions. Despite these imperfections, the experimentally observed success probability exceeds the $50\%$ limit of traditional fusion operations and significantly surpasses the percolation threshold ($58.98\%$) by 17 standard deviations, demonstrating the potential of constructing a larger graph-state using a boosted type-$\mathrm{\uppercase\expandafter{\romannumeral2}}$ fusion gate.

We further evaluate the validity of our fusion gate by measuring the fidelity of the final generated two-photon state $\rho_{14}$. For simplicity, we perform a joint projective measurement of the Bell state $|\Psi^-\rangle_{14}$, which corresponds to the post-selected outcome $|\Psi^-\rangle_{23}$ obtained through the Bell-state measurement. Two-photon correlation measurements are conducted along the basis $(+-, RL, HV)$, where $|\pm\rangle=(|H\rangle\pm|V\rangle)/\sqrt2$ and $|L/R\rangle=(|H\rangle)\pm i|V\rangle/\sqrt2$. The data are plotted in Fig.~\ref{fig4}(b). We calculate the state fidelity using $F=\text{Tr}(\rho_{exp}|\Psi^-\rangle_{14}\langle\Psi^-|_{14})=67(2)\%$. The measured fidelity exceeds the classical threshold of 0.5 by more than 8 standard deviations, providing strong evidence of genuine entanglement between photon pairs 1 and 4 and demonstrating the effectiveness of the fusion operations. Notably, this marks the first direct verification of the effectiveness of type-$\mathrm{\uppercase\expandafter{\romannumeral2}}$ fusion gates through observed entanglement between photons.


In summary, we have experimentally demonstrated the feasibility of scalable linear quantum computing by implementing fusion gates with a success probability of 71.0(7)\%, surpassing the percolation threshold necessary for constructing large graph-states from 3-photon GHZ states. Our simulations further indicate that our fusion gate exponentially expands the connected scale of the graph-state compared to conventional approaches. Our approach utilizes high-quality single photons generated from a solid-state system, ensuring both high efficiency and fidelity in our eight-photon interference experiment. Additionally, auxiliary states can be deterministically generated via HOM interference, significantly bolstering the practicality of our method.

Recent advancements in graph-state photon sources, as highlighted in Refs.~\cite{chen2023heralded,cao2023photonic,meng_2024_deterministic,thomas_2022_efficient}, have demonstrated their potential for scaling up graph-states. Larger graph-states could be constructed by multiplexing single entangled photon sources or integrating multiple sources on chips. Building upon these advancements, our work provides a foundational step toward scalable linear optical quantum computing.

\bibliography{sn-bibliography}

\end{document}